\documentclass{aa}

\usepackage{graphicx}
\usepackage{xcolor}
\usepackage{txfonts}
\usepackage[draft]{hyperref}
\newcommand{\nmem}{405}  
\newcommand{\nper}{129} 
\newcommand{\nmemlc}{222}  
\newcommand{\ngoodlc}{179}	
\newcommand{\ngoodper}{126} 

\begin{document}

   \title{A rotational age for the open cluster NGC\,2281 \thanks{The full Tables 1 and 2 are only available in electronic form at the CDS via anonymous ftp to cdsarc.u-strasbg.fr (130.79.128.5) or via http://cdsweb.u-strasbg.fr/cgi-bin/qcat?J/A+A/}}

   \subtitle{The possibility of using the fast rotator sequence for accurate cluster age determinations}

   \author{D. J. Fritzewski
          \inst{1, 2}
          \and
          S. A. Barnes
          \inst{1,3}
          \and
          J. Weingrill
          \inst{1}
          \and
      	  T. Granzer
      	  \inst{1}
      	  \and
      	  E. Cole-Kodikara
      	  \inst{1}
      	  \and
      	  K. G. Strassmeier
      	  \inst{1,4}
          }

   \institute{Leibniz Institute for Astrophysics Potsdam (AIP), An der Sternwarte 	16, 14482 Potsdam, Germany
   	\and
    Institute of Astronomy, KU Leuven, Celestijnenlaan 200D, 3001, Leuven, Belgium\\
    \email{dario.fritzewski@kuleuven.be}
    \and
   	Space Science Institute, 4750 Walnut St., Boulder, CO 80301, USA
    \and
    Institut f\"ur Physik und Astronomie, Universit\"at Potsdam, Karl-Liebknecht-Straße 24/25, 14476 Potsdam, Germany
   }

   \date{}


  \abstract
   {Cool star rotation periods have become an important tool in determining ages of open clusters.}
   {We aim to estimate the age of the open cluster NGC\,2281 based on the rotational properties of its low-mass members. Previous age estimates for this open cluster range from 275\,Myr to 630\,Myr.}
   {Based on an eight month-long photometric time series obtained at the 1.2\,m robotic STELLA telescope in Tenerife, we measured rotation periods for \ngoodper{} cool star members (70\,\% of the observed members) of NGC\,2281. }
   {The large set of rotation periods allows us to construct a rich colour-period diagram for NGC\,2281 with very few outliers above the slow rotator sequence. We identify an evolved fast rotator sequence which can be used to accurately age date the open cluster relative to other open clusters. Comparisons with M\,37 and M\,48 show that all three open clusters are roughly coeval, and we estimate the age of NGC\,2281 to be $435\pm50$\,Myr. Through comparisons with the younger NGC\,3532 and the older Praesepe, we determine the spin down rates of mid-K and early-M fast rotators to be significantly lower than for early-K stars. We suspect that the spin down of early-K fast rotators might be governed by an additional mass dependence.
    }
   {Finally, we show the path towards an empirical description of the evolved fast rotator sequences in open clusters.}


   \maketitle
%

\section{Introduction}

Open clusters are ideal calibration targets for stellar astrophysics as all member stars have shared properties such as metallicity and age. An age-ranked sequence of open clusters is of particular importance in understanding stellar evolution. However, accurate ages are not easily obtained, often due to the sparseness of higher-mass stars, increasing the uncertainties in fitting isochrones to colour-magnitude diagrams (CMD).

An alternative path to obtaining open cluster ages is through their cool main sequence stars, which are abundant in open clusters. These stars lose angular momentum in a mass-dependent way through the interaction of their magnetized wind with their magnetic field \citep{1958ApJ...128..664P, 1967ApJ...148..217W}. Solar-type stars approximately follow the well-known relationship between the equatorial rotational velocity and age, $t$: $V_\mathrm{eq}\sim t^{-0.5}$, first identified by \cite{1972ApJ...171..565S}. This has since been extended to stars of non-solar mass using less ambiguous rotation periods, leading to a procedure called gyrochronology, enabling the age $t=t(P, M)$ of a star to be derived \citep{2003ApJ...586..464B,2010ApJ...722..222B}.

Young open clusters host not only slow rotators that obey the famous Skumanich relation but also fast rotators \citep[e.g.][]{1982Msngr..28...15V, 1981IBVS.1957....1A}. Recently, we showed that the zero-age main sequence (ZAMS) distribution is essentially universal to within current data constraints (with the proviso about metallicity, \citealt{2020A&A...641A..51F}). It is bimodal in period distribution. Clusters of Hyades age have unimodal rotation period distributions, at least among FGK stars, while clusters of younger ages host stars in transition from fast to slow rotation in a mass-ranked sequence \citep{2021A&A...652A..60F}. It is important to understand this transition from bimodality (or at least a wide dispersion in rotation periods) to unimodality. Despite the obvious observational signatures, a clear picture of the physical reasoning behind the transition has yet to be found. In spite of this, the clear mass dependence might make it possible to construct an empirical age estimator based on the stars in transition from fast to slow rotation \citep{2003ApJ...586..464B}.

Comparing the entire rotation period distributions of open clusters of known ages purely observationally to those of unknown or insecure ages is a powerful tool for obtaining cluster age estimates. It has allowed \cite{2019AJ....158...77C} to show that the Meingast\,1 \citep{2019A&A...622L..13M,2020A&A...639A..64R} stellar stream is significantly younger than its isochronal age determination suggested because its stars rotate as fast as stars in the Pleaides. Similarly, \cite{2021AJ....162..197B} used rotation periods in concert with lithium abundance measurements to prove that the halo of the open cluster NGC\,2516 has the same age as the core of the cluster. For the stellar stream Theia 456 \citep{2019AJ....158..122K}, rotation confirmed the isochronal age \citep{2022AJ....163..275A}. Stellar rotation was also used to refine the age estimate of the nearby moving group Group~X \citep{2017AJ....153..257O,2018ApJ...863...91F,2018ApJ...862..106T} and to show that it is coeval with the open cluster NGC\,3532 \citep{2022A&A...657L...3M, 2022AJ....164..115N}.

The distribution of nearby, and therefore accessible, open clusters is sharply peaked near ZAMS ages, and then declines rapidly with increasing cluster age.
Rich clusters within about 500\,pc are rare, and the proximity of Ru\,147 (2.5\,Gyr, \citealt{2020A&A...644A..16G}) and M\,67 (4\,Gyr, \citealt{2016ApJ...823...16B}) is fortuitous, enabling rotation period studies to extend almost to solar age.
The younger objects such as the Pleiades (ZAMS age), Hyades, and Praesepe (both $\sim${}600\,Myr) have been studied repeatedly using ground- and space-based observatories \citep{1984ApJ...280..202S, 1987ApJ...321..459R, 1993PASP..105.1407P, 1995PASP..107..211P, 2004A&A...421..259S, 2010MNRAS.408..475H, 2011ApJ...740..110A, 2011MNRAS.413.2218D, 2011MNRAS.413.2595S, 2014MNRAS.442.2081K, 2014ApJ...795..161D, 2016ApJ...822...47D, 2017ApJ...842...83D, 2019ApJ...879..100D, 2017ApJ...839...92R,  2021ApJ...921..167R}. However, clusters with intermediate ages are few in number, and fail to simultaneously satisfy our criteria of richness and proximity. M\,34 ($\sim${}220\,Myr) went partway towards this goal \citep{2011ApJ...733..115M}, but its age was not significantly beyond the ZAMS. M\,37 ($\sim${}$ 450-550$\,Myr) is also rich but distant, requiring photometry from a 6\,m telescope \citep{2009ApJ...691..342H}. M\,48 ($\sim{}$450\,Myr) was observed by \cite{2015A&A...583A..73B}, but is not especially rich.
We recently studied NGC\,3532 ($\sim${}300\,Myr, \citealt{2021A&A...652A..60F}), which is unique in richness, age, and proximity.

In this work, we investigate the open cluster \object{NGC\,2281} which is also proximate (just beyond the nominal 500\,pc horizon, at 525\,pc; \citealt{2021A&A...649A...2L}) and whose age estimates span a large range with a bimodal distribution. \cite{1968ArA.....5....1L} and \cite{1976ApJS...30..451H} found relatively young ages for the cluster of 250\,Myr and 280\,Myr, respectively. \cite{1981A&A....97..235M} estimated 300\,Myr and placed it 40\,Myr older then the rich southern open cluster NGC\,3532. More recent work finds the cluster to be twice as old. \cite{2005A&A...438.1163K} found a cluster age of 610\,Myr. \cite{2019AJ....158...35S} use near-ultraviolet observations to obtain an age of 630\,Myr, similar to the Hyades. However, this estimate is based on a larger reddening. Based on \emph{Gaia} data, \cite{2019A&A...623A.108B} place the age at 600\,Myr, while \cite{2019AJ....158..122K} find a much younger age of 275\,Myr. By observing chromospheric activity \cite{2019ApJ...887...84Z} obtain two ages of 526\,Myr (from \ion{Ca}{II}) and 566\,Myr (from H$\alpha$). \cite{2021Galax...9....7T} found 630\,Myr by assuming an extinction value of $E(B-V)=0.123$, close to the typically used $E(B-V)=0.1$. As rotation periods of young open clusters have previously been shown to be able to sensitively date open clusters and stellar streams, we aim to place similar constraints on the age of NGC\,2281 with their help. Obtaining a well-constrained cluster age will also allow us to probe the rotational transition and examine whether the universality seen among ZAMS clusters also is obtained for somewhat older clusters.

Early work on NGC\,2281 includes a proper motion study of the field \citep{1959AJ.....64..170V}, a few photometric surveys \citep{1938AnHar.106...39C,1961ApJ...134..602P, 1964AJ.....69R.529A,1978NInfo..41..101B,1978PASJ...30..123Y}, spectroscopy of its giant member \citep{1964ApJ...140..858W}, and observations of Ap stars in the cluster \citep{1977ApJ...212..723H}. An early photometric survey found a reddening towards the cluster of $E(B-V)=0.1$ \citep{1961ApJ...134..602P} which was confirmed by \cite{1981A&A...104..185N} ($E(B-V)=0.12$) and \cite{1987PASP...99.1089G} ($E(B-V)=0.11$ from spectroscopy). More dedicated studies of NGC\,2281 have only recently been conducted while earlier work included typically only a few stars of the cluster to measure certain properties such as metallicity.

Recently, \cite{2018PASP..130h4206S} presented far-ultraviolet observations and the precise data provided by the \emph{Gaia} mission enable detailed membership analysis \citep{2019PASJ...71...62G} which \cite{2021Galax...9....7T} combined with ground-based photometry. In our work, we add to recent developments, and derive rotation periods from long-term time series observations of NGC\,2281.

The \emph{Gaia} revolution has also allowed for membership studies of open clusters on the Galactic scale including thousands of clusters (e.g. \citealt{2018A&A...618A..93C, 2019AJ....158..122K}). For this work, we use the membership probabilities of \cite{2019AJ....158..122K}. NGC\,2281 was also included in LAMOST \citep{2012RAA....12.1197C} from which \cite{2019ApJ...887...84Z} investigated the chromospheric activity-age relation, \cite{2020ApJ...903...93N} estimated the binarity fraction in open clusters, and \cite{2021ApJ...908..207Z} measured main sequence ages of a large stellar sample including NGC\,2281 as a calibrator.

Several metallicity estimates from both photometry and spectroscopy exist for NGC\,2281 which mostly agree and find a slightly sub-Solar metallicity. \cite{1985A&A...147...39C} obtained a photometric metallicity of [Fe/H]$=-0.07$ and \cite{2017MNRAS.469.3042N} found [Fe/H]$=-0.03$ also from photometry. From the LAMOST spectroscopic survey, \cite{2019ApJ...887...84Z} derived [Fe/H]$=-0.03$ and \cite{2020ApJ...903...93N} measured [Fe/H]$=-0.1$.

The paper is structured as follows. In Sect.~\ref{sec:obs}, we present our observations and photometry leading to the creation of light curves. After defining the membership in Sect.~\ref{sec:members}, we analyse the light curves in Sect.~\ref{sec:timeseries} and obtain stellar rotation periods. In Sect.~\ref{sec:activity}, we analyse the photometric variability of the cluster members. Thereafter, we present the colour-period diagram (CPD) in Sect.~\ref{sec:CPD} and estimate the rotational age of NGC\,2281 in Sect.~\ref{sec:age} through comparisons with other open clusters. In Sect.~\ref{sec:binout}, we take a closer look at the influence of binarity on rotation and outliers to the sequences in the CPD. Finally, we estimate the rotational age of NGC\,2281  and discuss implications for the spin down of fast rotators in Sect.~\ref{sec:spindown}.

\section{Observations and photometry}
\label{sec:obs}
\subsection{Observations and data reduction}

We observed NGC\,2281 in three campaigns over 220 nights in 2013, 2016/2017, and 2018 with the robotic 1.2\,m STELLA observatory \citep{2004AN....325..513G, 2004AN....325..527S, 2010AdAst2010E..19S, 2016SPIE.9910E..0NW} on the Izana ridge at the IAC site in Tenerife. The majority of the observations were obtained in the 2016/2017 campaign and after the quality cuts explained below only a few good frames were left from the other two observing seasons.
Hence, we decided to concentrate our analysis on the largest campaign as we expect no additional scientific results from adding in the short time series. Our data set spans
158 nights within the 246 days between 21 August 2016 and 24 April 2017.

The robotic STELLA observatory consists of twin telescopes  with 1.2\,m mirrors, one of which is dedicated to photometry and the other to spectroscopy. This study utilizes the
photometric telescope STELLA-I which is equipped with the Wide-field STELLA Imaging Photometer (WiFSIP). The detector has a resolution of $4064\times 4064$\,pixels and is read out with four amplifiers. Its usable field of view is $22\arcmin \times 22\arcmin{}$ which results in an 0.32\arcsec\,px$^{-1}$ pixel scale.

All frames obtained with WiFSIP are reduced in the same, standard pipeline \citep{2016SPIE.9910E..0NW}. After amplifier-crosstalk correction, overscan and bias subtractions, the frames are flat-fielded using twilight flats. Flat-field corrected frames are processed with \texttt{SExtractor} and matched with \texttt{WCSTools} to obtain a WCS solution.

The data from STELLA/WiFSIP are not uniform in quality, mostly due the fact that the telescope operates in almost all weather conditions that are not actually dangerous to the telescope. In particular, observations are also carried out on non-photometric nights and during nights with bad seeing and/or high winds. Depending on the science project, this can lead to a significant fraction of the images having to be excluded from the analysis.

We used several criteria to sort our images between usable and unusable for extracting photometry.
We excluded all images with a FWHM larger than 4.5\arcsec, as we found these images typically to be streaks rather than circular. With the same reasoning, we excluded stars with an ellipticity $>0.3$. Some images suffer from a high and non-uniform background due to scattered moon light.  Images with a background significantly higher than the median of moonless nights were rejected. Surprisingly, we also had to exclude images with seeing well below the average seeing (better than 1.8\arcsec) because these sharp images could not be reduced in concert with the other images with wider point spread functions (psf). Finally, a few frames had to be removed due to technical difficulties such as a warm image sensor or a (partially) closed dome. However, these are just a few percent of all rejected images.

\begin{figure}
	\includegraphics[width=\columnwidth]{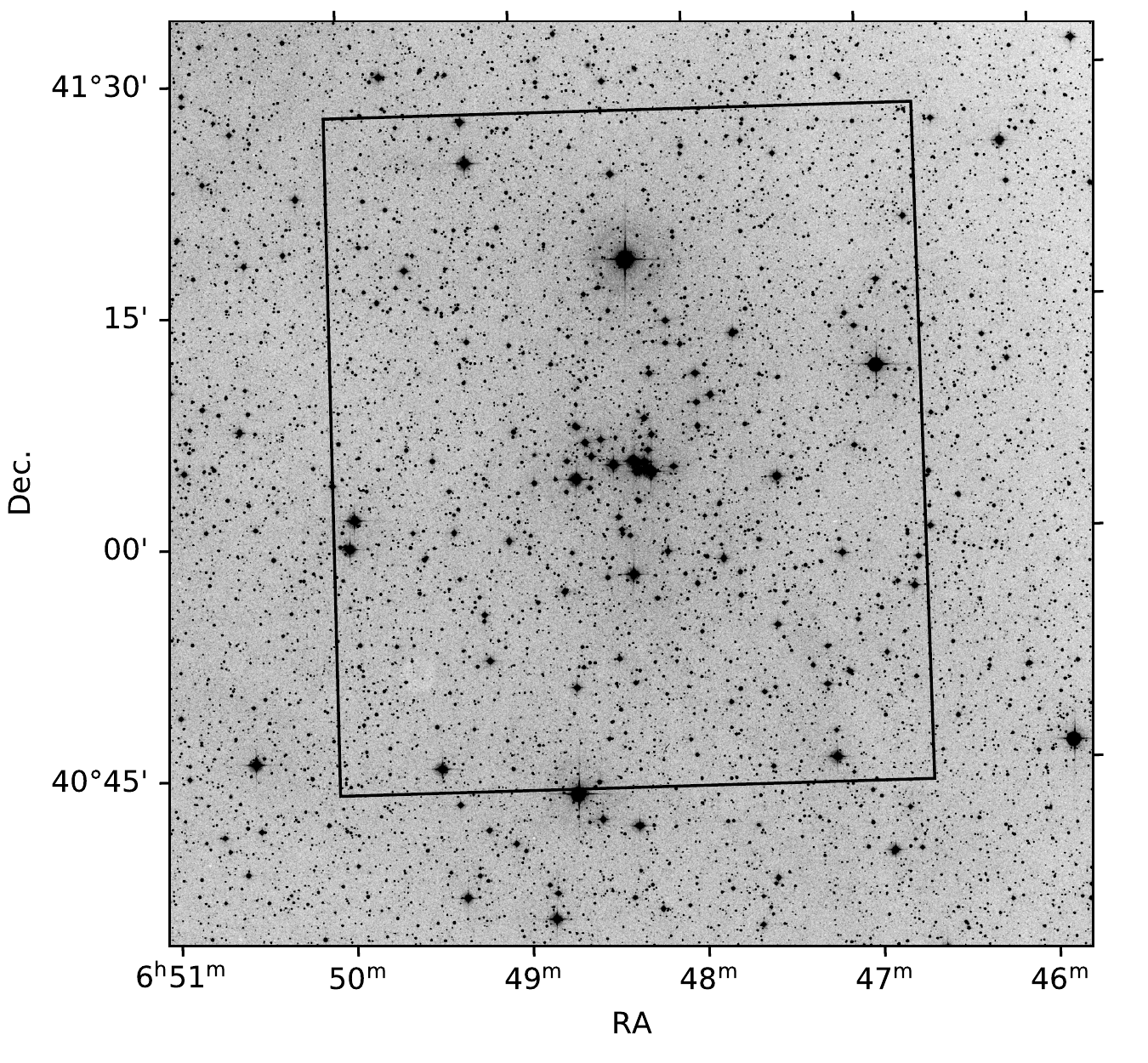}
	\caption{Sky position of the combined field superimposed on a DSS image. The observations were centred on the core of the open cluster. The observed area is not exactly square because the overlap between the individual fields is larger in right ascension than in declination.}
	\label{fig:mosaic}
\end{figure}

Overall, 5\,377 images were obtained in the observing campaign of which 2\,047 had to be discarded due to the above-mentioned reasons\footnote{The large number of discarded frames is a generic feature of observing with robotic telescopes. They observe in conditions in which a human observer would probably not, and where the selection of good observations take place even before observing. However, with the robotic telescope approach, the duty cycle of good frames is typically better.}. Broken down by individual reasons, we find 958 images being removed due to a high background, 1147 due to high ellipticity, 1144 due to a large FWHM, 375 due to a small FWHM, but only 48 due to a warm sensor and 28 due to problems with the dome\footnote{These numbers do not add up to 2\,047 because often multiple conditions are met.}. For the time series photometry, we used 3\,330 of the images. They are distributed over four fields which cover the central parts of the open cluster (Fig.~\ref{fig:mosaic}). For each field, we obtained observations with
exposure times of 24\,s and 120\,s in the $V$ filter and 600\,s in $R_c$.

\subsection{Photometry with \texttt{Daophot II}}
For each field and exposure setting, we performed psf-photometry with \texttt{Daophot II} and then constructed light curves with \texttt{Allframe}  \citep{1987PASP...99..191S, 1994PASP..106..250S, 2003PASP..115..413S}. In detail, the workflow was as follows. In each image, all possible sources were identified and preliminary aperture photometry was taken to identify bad detections such as hot pixels. Afterwards, we matched each image against the reference image for which we chose nights with a consistently low background, a medium FWHM, and small ellipticity. However due to the ever-changing image quality, we could not use the same night for all exposure series. Instead, we picked the reference frame for each of the twelve series (four fields times three exposure settings) independently. In each reference image, we selected about 180 stars which are used to fit the psf. The same set of stars was identified in all images.

With our psf stars at hand, we fitted a Lorentz function\footnote{Analytic PSF = 4 in \texttt{Daophot II}} as the psf for each image. Consecutively, we used the programme \texttt{Allstar} to subtract the stellar flux from each image. Afterwards, we matched the stars' positions in the different images to within 15\,px with \texttt{Daomaster}\footnote{Due to the wide and sometimes elongated psf in the images, we did not match to a closer distance because this would have rejected too many stars.}, created homogenized photometry, and constructed light curves with \texttt{Allframe}. Further details on our \texttt{Daophot} workflow can be found in \cite{2020A&A...641A..51F}.

\section{Membership and binarity}
\label{sec:members}
\subsection{Membership}
For our analysis of variable stars in the field of NGC\,2281, we concentrated on the cluster members. Various membership analyses of NGC\,2281 based on \emph{Gaia} data are available. We chose \cite{2019AJ....158..122K} as our primary membership list because the authors included not only the cluster core but also extended regions of the cluster in their analysis. However, \cite{2019AJ....158..122K} made a quality cut in the \emph{Gaia} data and did not include stars fainter than $G=18$\,mag. Our photometry, on the other hand, includes stars fainter than this and we even possess light curves of stars beyond the \emph{Gaia} faint limit. Hence, we enriched the membership list of \cite{2019AJ....158..122K} with additional probable members from our own analysis.

For the additional members, we used proper motions and photometry from \emph{Gaia} DR3 \citep{2021A&A...649A...3R, 2021A&A...649A...2L} without any quality cuts to gain an inclusive set of cluster members\footnote{The rotational analysis later acts as an additional membership criterion excluding older field stars with a low activity level.}. In addition to the \emph{Gaia} data, we used infrared photometry from ALLWISE \citep{2010AJ....140.1868W, 2011ApJ...731...53M}.

We classified all stars with a proper motion within 0.75\,mas\,yr$^{-1}$ of the cluster mean ($\mu_{\alpha\star} = -2.952$, $\mu_{\alpha\star}\delta = -8.245$) as proper motion members. Photometric members were determined based on their distance to the empirical colour-magnitude sequence \citep{2013ApJS..208....9P}\footnote{\url{http://www.pas.rochester.edu/~emamajek/EEM\_dwarf\_UBVIJHK\_colors\_Teff.txt}} in a [$(G-G_\mathrm{RP})_0$, $G_0$] CMD and [$(G-G_\mathrm{RP})_0$, $(W1-W2)_0$] colour-colour diagram (more information can be found in the Appendix~\ref{app:members}).

We combined both criteria and defined the additional members as stars which are both proper motion and photometric members. In cases where no proper motion is available the additional members are based purely on photometry. As a consequence, the single star cluster sequence, as shown in Fig.~\ref{fig:members}, widens for the lowest mass stars and photometric binaries in this faint region are not considered. As expected, our additional members can mostly be found among the lower-mass stars which were not included in the membership analysis of \cite{2019AJ....158..122K} due to their faintness.

NGC\,2281 is not a well studied open cluster and no dedicated radial velocity study is available. However, \emph{Gaia}~DR3 provides a significant number of radial velocity measurements \citep{2022arXiv220605902K}. As the precision of these radial velocities is very variable, we only removed obvious radial velocity non-members with $\Delta v_\mathrm{r}>10$\,km\,s$^{-1}$ relative to the mean velocity of the cluster $v_\mathrm{r}=20$\,km\,s$^{-1}$. We find 20 stars included above that are radial velocity non-members.

In total,we find \nmem{} members in the field of NGC\,2281 (see Table~\ref{tab:members}). Of these, 260 are in common with the analysis by \cite{2019AJ....158..122K} while the remaining 145 members have mostly $G>18$\,mag and were therefore not considered in \cite{2019AJ....158..122K}. We conducted our membership analysis in an area somewhat larger than our photometric coverage and we possess light curves for \nmemlc{} of the members.

\begin{table*}
    \caption{Membership list for NGC\,2281. The full table is available online.}
    \label{tab:members}
    \begin{tabular}{rlrrrrlr}
        \hline
        \hline
        ID & Designation & RA & Dec. & $G$ & $(G-G_\mathrm{RP})_0$& KC19 member & $R_\mathrm{var,V120}$\\
        & & (\degr) & (\degr) & (mag) & (mag) & & (mag)\\
        \hline
        96 & Gaia DR3 945455617635072000 & 102.17281 & 40.72973 & 12.40 & 0.350 & y & 0.028\\
        100 & Gaia DR3 945464933420822912 & 101.70701 & 40.73059 & 17.25 & 0.939 & y & 0.095\\
        114 & Gaia DR3 951457306152396544 & 102.26910 & 40.73343 & 15.34 & 0.675 & y & 0.037\\
        162 & Gaia DR3 945467304242755072 & 101.95399 & 40.73987 & 14.86 & 0.716 & y & 0.055\\
        178 & Gaia DR3 945465139579247872 & 101.78642 & 40.74175 & 12.82 & 0.407 & y & 0.029\\
        252 & Gaia DR3 945456584004370816 & 102.10366 & 40.74895 & 16.69 & 0.910 & y & 0.063\\
        258 & Gaia DR3 945465242658462080 & 101.79579 & 40.74985 & 15.68 & 0.734 & y & 0.039\\
        317 & Gaia DR3 951457787188730240 & 102.27181 & 40.75626 & 13.03 & 0.407 & y & 0.030\\
        329 & Gaia DR3 951411603403226496 & 102.44962 & 40.75827 & 19.38 & 1.144 & n & \dots\\

        \dots&&&&&&&\\
        \hline
    \end{tabular}
\tablefoot{\emph{ID} gives our identification number, \emph{KC19 member} idicates whether the star is in \cite{2019AJ....158..122K}, $R_\mathrm{var,V120}$ is the variability amplitude as measured from the V120 light curves. The remainig columns provide \emph{Gaia} DR3 data.}
\end{table*}

\subsection{Binarity}
\begin{figure}
    \includegraphics[width=\columnwidth]{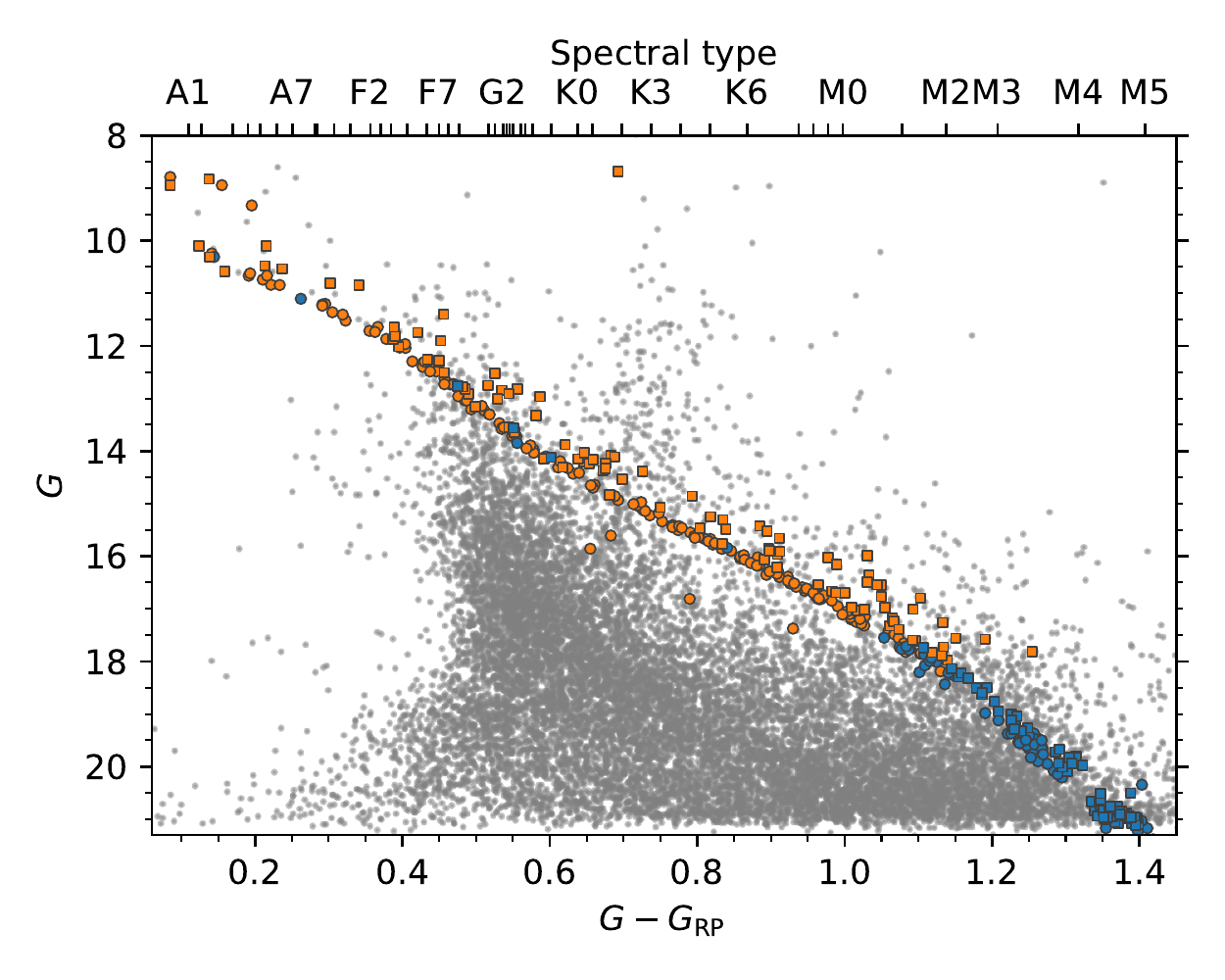}
    \caption{Colour-magnitude diagram of the field of NGC\,2281. Cluster members are marked with larger symbols, with orange we denote members from \cite{2019AJ....158..122K} and our additional members are shown in blue. Potential binaries are marked with squares. The top axis marks the spectral type for cluster members and is based on \cite{2013ApJS..208....9P}.}
    \label{fig:members}
\end{figure}

Stellar companions can influence the rotational evolution. Therefore, we identified potential binary stars among the members with three criteria. Firstly, we marked all stars 0.25\,mag above the manually traced main sequence ridge line in the \emph{Gaia} CMD (Fig.~\ref{fig:members}) as potential photometric binaries. The accurate \emph{Gaia} photometry allows for such low magnitude differences to be reliably identified and enables us to classify non-equal-mass binaries as potential binaries. Secondly, we used the reduced unit weight error (RUWE) from \emph{Gaia} DR3 \citep{2021A&A...649A...2L} to identify additional non-photometric binaries. RUWE is a measure of the astrometric noise and a high RUWE value indicates that the \emph{Gaia} pipeline was not able to locate the position of the star accurately. Among others, this can be caused by photo-centre shifts due to an unresolved companion. Because other factors could also lead to high RUWE values, we call the selected stars potential binaries. We applied a threshold of $\text{RUWE}>1.2$ \citep{2020MNRAS.496.1922B} for them. Thirdly, \emph{Gaia} also provides radial velocity amplitudes for some objects. As all of the stars with amplitudes are variable in radial velocity, we also include them as binaries here. In the CMD in Fig.~\ref{fig:members} and in all subsequent relevant figures the potential binaries are marked with squares. We do not distinguish between photometric, RUWE, or radial velocity binaries in the figures.

\section{Time series analysis}
\label{sec:timeseries}
For all light curves produced by \texttt{Daophot II} (Sect.~\ref{sec:obs}), we compute periodograms with four different methods. As in our prior work (e.g. \citealt{2020A&A...641A..51F}), we use multiple periodograms of the same light curve to correctly identify the rotation period and estimate its uncertainty.

For this work, we applied the generalized Lomb-Scargle periodogram \citep{2009A&A...496..577Z}, the \texttt{clean} periodogram \citep{1987AJ.....93..968R, 2001SoPh..203..381C}, phase-disperion minimization \citep{1978ApJ...224..953S}, and the string length method \citep{1983MNRAS.203..917D}. The former two methods are Fourier-based while the latter two employ phase-folding to find the optimal period. We found these methods to be reliable in our previous work and did not explore further options as they would not add information.

\begin{figure*}
    \centering
    \includegraphics[width=\textwidth]{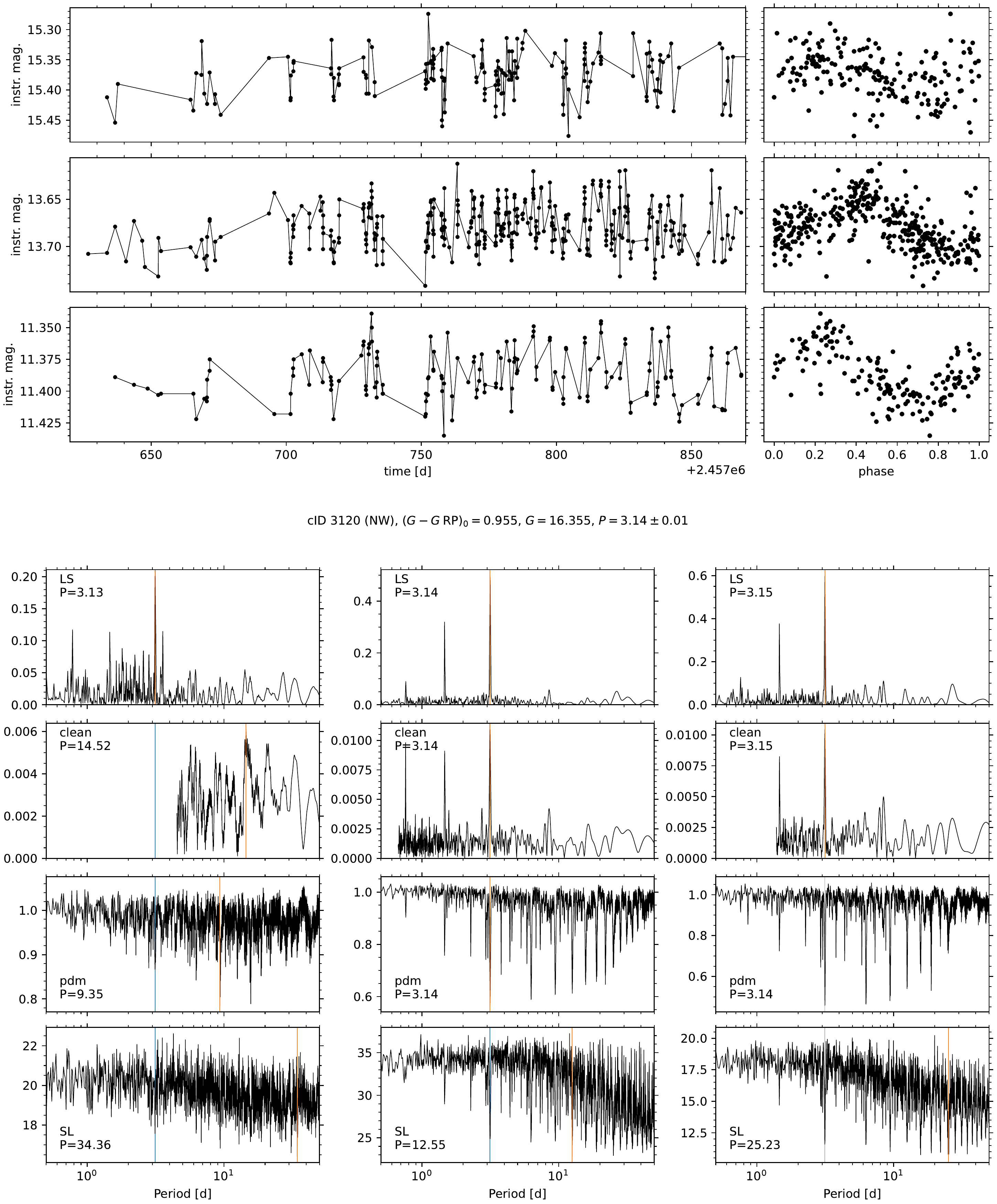}
    \caption{Example of our overview diagram as produced for every star. The \emph{top} panels show the full (\emph{left}) and phase-folded light curves (\emph{right}) for each exposure setting. From top to bottom, we show the $V$ 24\,s, $V$ 120\,s, $I_c$ 600\,s. At the \emph{bottom}, we display the four periodograms for each exposure setting, now from left to right. In each periodogram, we mark the best period with an orange line and give its value at the left. The finally chosen period is marked with the blue line and noted in the text in the centre. We note that with the joint analysis, we are able to identify the true rotation period even in periodograms with higher alias peaks. The width of the blue line indicates the period uncertainty.}
    \label{fig:overview}
\end{figure*}

Although computed for all observed stars, we concentrate our periodogram analysis on the members of NGC\,2281 as determined above. Analysing each of the four quadrants separately, we evaluate all members with light curves within the quadrant at the three exposure times by using the four methods listed above and create an overview of twelve periodograms, if available\footnote{Rotation periods of members in multiple fields are combined later, adding additional confidence to the period.}. One example is shown in Figure~\ref{fig:overview}. These diagrams were evaluated manually simultaneously to identify a common peak in the periodograms: the rotation period.

We focused on three main criteria to identify rotation periods with the periodograms. Firstly, any period should be recognisable in the periodograms and associated with a strong peak or dip (depending on the periodogram type). Secondly, the approximate position of these peaks has to be  consistent across the different periodograms and across the different light curves. Finally, the chosen period must not be a common alias period which is found in many light curves. For our light curves, we identified $P_\mathrm{alias} \approx [$0.5\,d{,} 1\,d{,} 12.5\,d{,} 25\,d] as alias periods. The former two originate in the daily cycle, while the latter two are likely caused by the lunar cycle. A high sky background is more likely to occur around full Moon. By removing frames near that time, the light curve contains stretches of observations slightly shorter than the lunar cycle.

In particular, the second constraint has helped us identifying potential rotation periods even in low signal-to-noise conditions. Due to the properties of the STELLA observations the window functions of the three independent light curves of each star are not identical, hence the only consistent peaks across all periodograms are either the most prominent alias periods or the true underlying signal.

After identifying the most probable rotation period, we read off the value manually from the periodograms and identify in each of the twelve periodograms  the highest peak within 15\,\% of the estimated period. By using the highest peak within this window, the measured rotation periods from each periodogram are not subject to any further manual biases which makes the methods repeatable even when a different period was read off from the periodograms. In our experience, a period can reliable be measured when the estimated period is within 0.25\,d of the true value for faster rotators and with 1\,d fast slower rotators. Cases in which the highest peak is different from the intended (because of multiple nearby peaks) are rare.

The final rotation period is the simple mean of all twelve (or more if the star is found in multiple fields) determined periods. The largest difference between this mean value and the individual periods is taken as the uncertainty. As shown in \cite{2020A&A...641A..51F}, this approach is both efficient in locating the correct period and in estimating the period uncertainty.

Light curves phase-folded with the estimated rotation period rarely show signs of consistent rotational signals. However, this is mainly due to the noisy data which is also expressed by the periodograms in showing the alias periods as the main periodic components. Some of the observed noise might be intrinsic to the stars. Given the long-term time series over eight months it is not surprising for star spots to disappear and to appear at different longitudes. This shifts the starspot-induced signal out of phase and make it disappear in phased light curves. We note that despite these drawbacks the resulting colour-period diagram in the next section is clean, has a shape similar to that of other cluster distributions, and few outliers.

In total, we found \nper{} rotation periods (Table~\ref{tab:periods}) among the \nmemlc{} members for which we constructed light curves. The members include stars of all masses, including A\,type stars at the turn-off. Rotation periods measured from star-spot induced variability can only be found among cool stars. When restricting the members to a range from late F ($G-G_\mathrm{RP}=0.4$) to early M ($G-G_\mathrm{RP}=1.25$, our detection limit), the sample includes \ngoodlc{} stars of which we found rotation periods for \ngoodper. In summary, we are able to identify 70\,\% of the members in this key mass range as periodic rotators.

\begin{table*}
    \caption{Rotation periods for members of NGC\,2281. The full table is available online.}
    \label{tab:periods}
    \begin{tabular}{rlrrrrrr}
        \hline
        \hline
        ID & Designation & RA & Dec. & $(G-G_\mathrm{RP})_0$&$P_\mathrm{rot}$ &$\Delta P_\mathrm{rot}$ & $Ro$\\
        & & (\degr) & (\degr) & (mag) & (d) & (d) & \\
        \hline
100 & Gaia DR3 945464933420822912 & 101.70701 & 40.73059 & 0.939 & 14.87 & 1.57 & 0.102\\
114 & Gaia DR3 951457306152396544 & 102.26910 & 40.73343 & 0.675 & 9.34 & 0.49 & 0.140\\
162 & Gaia DR3 945467304242755072 & 101.95399 & 40.73987 & 0.716 & 9.29 & 0.20 & 0.123\\
178 & Gaia DR3 945465139579247872 & 101.78642 & 40.74175 & 0.407 & 2.91 & 0.17 & 0.140\\
252 & Gaia DR3 945456584004370816 & 102.10366 & 40.74895 & 0.910 & 14.17 & 1.17 & 0.106\\
258 & Gaia DR3 945465242658462080 & 101.79579 & 40.74985 & 0.734 & 11.09 & 0.12 & 0.139\\
317 & Gaia DR3 951457787188730240 & 102.27181 & 40.75626 & 0.407 & 4.71 & 0.22 & 0.227\\
414 & Gaia DR3 945468025797248128 & 102.07856 & 40.76946 & 0.795 & 11.92 & 1.34 & 0.131\\
508 & Gaia DR3 951461291881895040 & 102.17473 & 40.78195 & 0.834 & 3.74 & 0.01 & 0.038\\

        \dots &&&&&&&\\
        \hline
    \end{tabular}
\tablefoot{$P_\mathrm{rot}$ and $\Delta P_\mathrm{rot}$ are the rotation period and its uncertainty, respectivily. $Ro$ gives the Rossby number.}
\end{table*}

\section{Photometric stellar activity}
\label{sec:activity}
\begin{figure*}
    \includegraphics[width=\textwidth]{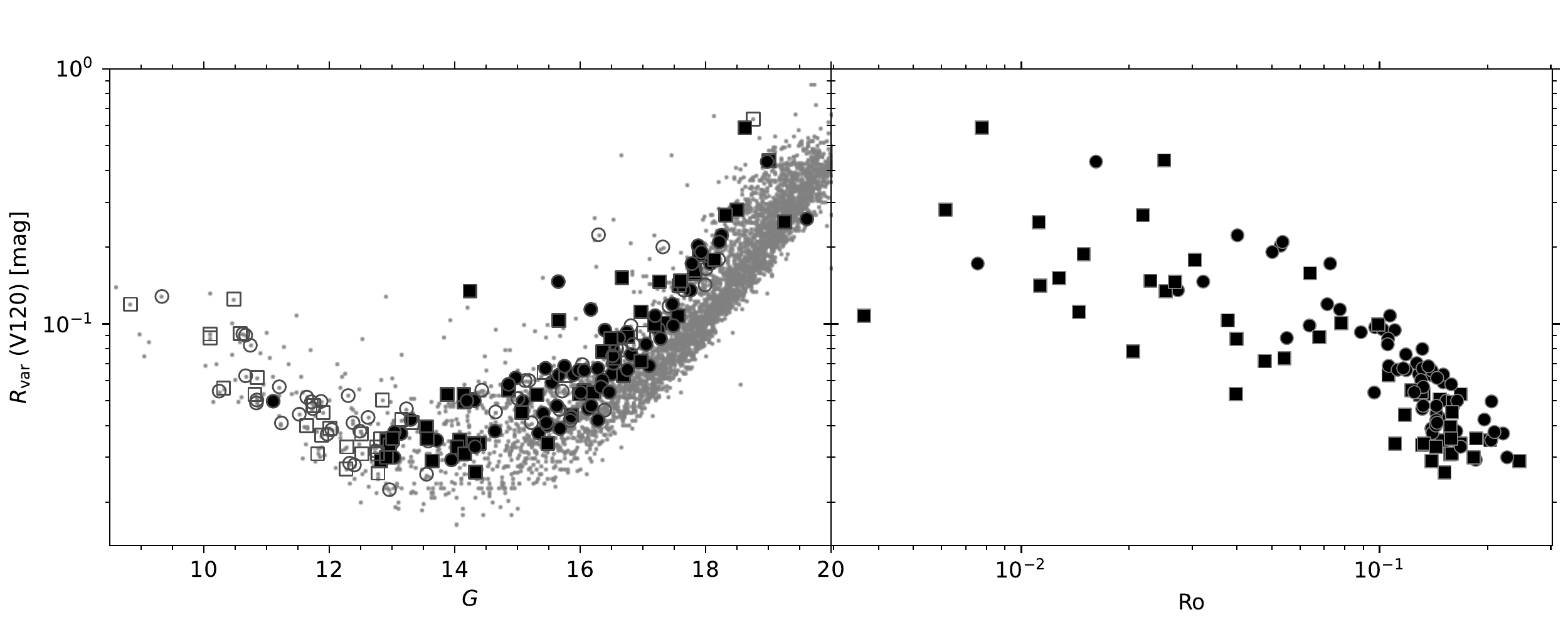}
    \caption{Photometric variability as a function of stellar properties. \emph{Left:} photometric variability, $R_\mathrm{var}$, against \emph{Gaia} $G$ magnitude based on light curve amplitudes in the V filter with 120\,s exposures. Rotators are shown with filled, other members of NGC\,2281 with unfilled symbols. Candidate binaries are marked with squares. Small grey symbols show the field stars. We note that all rotators are well above the noise level. \emph{Right:} corresponding rotation-variability diagram showing the rotators from the \emph{left} panel. Saturated fast rotators are found among the large amplitude stars, whereas slower unsaturated rotators form a sequence of decreasing activity with increasing Rossby number.}
    \label{fig:activity}
\end{figure*}

Before analysing the rotation periods derived above, we explore the stellar activity of members of NGC\,2281 because all magnetic activity phenomena are closely related to stellar rotation. An activity measure easily derivable from the light curves is the photospheric activity which manifests itself in the light curve amplitude. The difference between the fifth and ninety-fifth percentile ($R_\mathrm{var}$) is routinely used as the variability amplitude \citep{2011AJ....141...20B} and excludes outlying data points.

In the left panel of Fig.~\ref{fig:activity}, we show the variability amplitudes, $R_\mathrm{var}$, (as measured from the 120\,s exposure in the $V$ filter)\footnote{Variability amplitudes measured in the other filter-exposure time configurations show the same correlations. We chose V120 here because it has the most data points.} for all members of NGC\,2281 (with V120 light curves) against their \emph{Gaia} $G$ magnitude. The cluster members are clearly more variable than field stars of a similar brightness. Hence, the observed variability amplitude is not caused by photometric errors. The few members without a measured rotation period among the cool stars follow the rotators well. These stars may simply have an unfavourable spot configuration that has a too low signal-to-noise ratio for our period analysis.

In most cases, cluster members well above the normal variability for a given magnitude show a very clear rotational signal. Hence, the above average variability is not a consequence of noise or stray light from a neighbouring star but is truly astrophysical. All of these stars have short rotation periods, making these high variability amplitudes plausible. The high amplitude stars are all probable members of the cluster and are close to the cluster mean parameters in parallax and proper motion. We note that the high variability is not obviously related to binarity as half of the outliers are probable single stars.

The right panel of Fig.~\ref{fig:activity} folds in the rotational properties of the stars to construct a rotation-activity diagram. We show the same variability amplitude against the Rossby number $Ro$. The Rossby number is a mass-normalized rotation period and is calculated as $Ro = P_\mathrm{rot}/\tau_c$ with $\tau_c$ the convective turn-over time. For our work, we use the convective turn-over time from \cite{2010ApJ...721..675B}\footnote{Other estimates of the convective turn-over time differ mostly by a constant factor (c.f. Fig~6 of \citealt{2011ApJ...741...54C}) and would therefore only shift the x-axis.} and estimate it through the stellar effective temperature which was calculated from $(G-G_\mathrm{RP})_0$ \citep{2021MNRAS.507.2684C}.

The rotation-activity diagram has the expected shape known from other open clusters in photospheric, chromospheric \citep{2021A&A...656A.103F}, and coronal activity \citep{2020A&A...641A..51F}. Faster rotators (low Rossby number) have a high, saturated activity level that is independent of the rotation rate. Stars with $Ro\gtrsim 0.1$ are unsaturated and their activity levels are consonant with their larger Rossby numbers. In summary, we find the rotation-activity relation of NGC\,2281 to follow the expected shape.

\section{Colour-period diagram}
\label{sec:CPD}
\begin{figure}
    \includegraphics[width=\columnwidth]{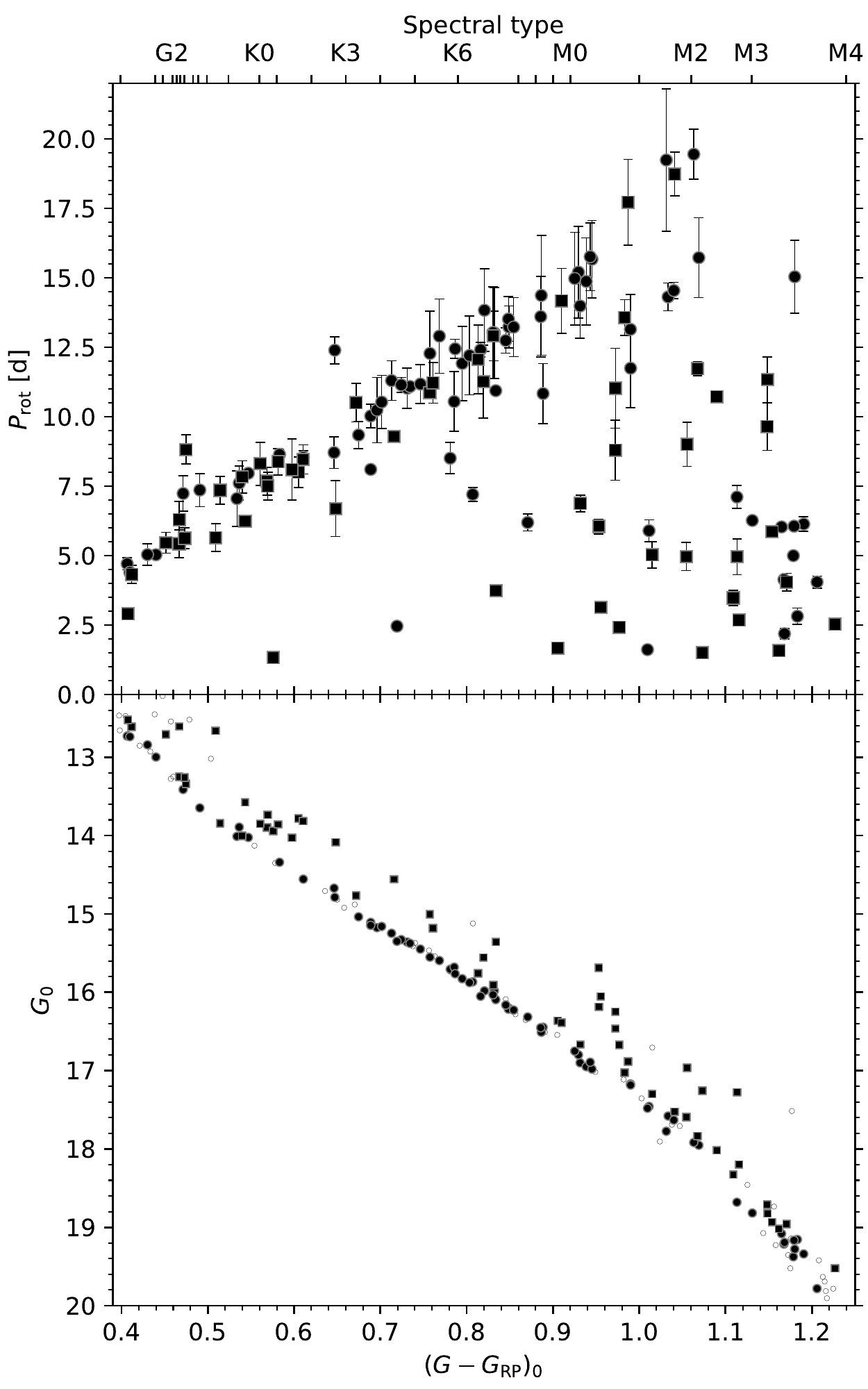}
    \caption{Detected rotators in the overall context of the open cluster. \emph{Top:} Colour-period diagram for members of NGC\,2281. It features a clear and well populated slow rotator sequence running diagonally from higher mass stars with periods $\lesssim 5$d up to $P_\mathrm{rot}\approx 20$\,d for the least massive stars in our sample. Only two outliers are evident above the sequence. Below this sequence, we find a few fast rotators, likely outliers, among the early K stars. Among the lower mass stars, we find several faster rotating stars providing evidence that this open cluster is certainly younger than the Hyades. Potential binaries are marked with squares. \emph{Bottom:} corresponding colour-magnitude diagram of the rotators (black) and non-rotators (small open circles).}
    \label{fig:cpd}
\end{figure}

Cool star rotation periods are best discussed in a mass-dependent way to understand their twofold dependence on mass and age. Our colour-period diagram (CPD) for NGC\,2281 in Fig.~\ref{fig:cpd} uses the \emph{Gaia}~DR3 colour $(G-G_\mathrm{RP})_0$ as a proxy for the stellar mass. With \nper{} rotation periods accounting for 70\,\% of the members in the cool star regime down to our detection limit the CPD is fairly complete.

The CPD for NGC\,2281 features a well-populated slow rotator sequence stretching from late-F with $P_\mathrm{rot}\approx 4$\,d to early-M stars with $P_\mathrm{rot}\approx 20$\,d with a few outliers. There are numerous faster rotating stars below this sequence, with the majority of them in the M\,dwarf regime. We also find some fast rotating outliers among the bluer stars which we subsequently discuss individually below after establishing the position of the fast rotator sequence. Among the reddest stars, the sample shows a wide spread in rotation periods as known from other young open clusters. An extended slow rotator sequence,with rotation periods extending to $\sim$30\,d (as found in NGC\,2516, \citealt{2020A&A...641A..51F}), is not observed here.
A dedicated search for longer periodic signals in the low-mass range was inconclusive, likely due to insufficient sensitivity at the faint end and a relatively strong 25\,d alias which disguises possible true period signals of similar length.

\section{Comparison with other open clusters and the rotational age of NGC\,2281}
\label{sec:age}

As shown in previous work (discussed, for instance, in the introduction), comparing the rotational distributions of different open clusters and stellar groups empirically is a powerful tool to estimate their age. In previous work on NGC\,2281, one group favoured an age similar to the age of NGC\,3532 (estimates in the range 250$-$300\,Myr) while another group favoured ages similar to that of the Hyades and Praesepe (526$-$630\,Myr, s. Tab.~\ref{tab:ages}). Fortunately, measured rotation periods are available for those clusters, as well as two of others of intermediate age, allowing us to rank the relevant clusters by rotational properties.

\begin{figure*}
    \includegraphics[width=\textwidth]{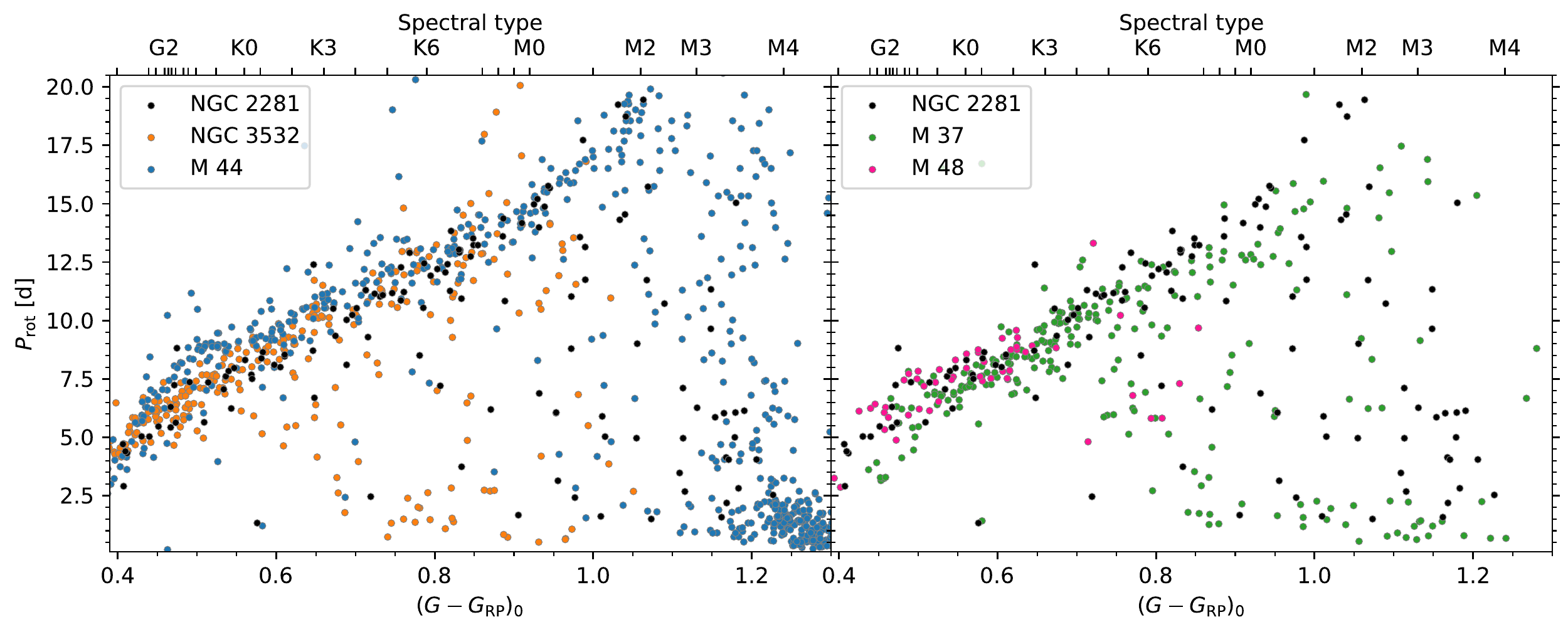}
    \caption{Comparison of the rotation period distribution of NGC\,2281 (black) with other open clusters of relevant age.
    \emph{Left:} NGC\,3532 (orange, 300\,Myr, \citealt{2021A&A...652A..60F}) and Praesepe (blue, 670\,Myr, \citealt{2019ApJ...879..100D}) bracket the previous age estimates. From the rotation periods it immediately follows that NGC\,2281 is younger than Praesepe and older than NGC\,3532, although perhaps closer in age to NGC\,3532 than to Praesepe.
    \emph{Right:} M\,37 (green, 450\,Myr, \citealt{2009ApJ...691..342H}) and M\,48 (pink, 420\,Myr, \citealt{2015A&A...583A..73B}) have nearly identical rotation period distributions to that of NGC\,2281 (black), including the faster rotating stars. This suggests similar ages for M\,37, M\,48 and NGC\,2281.
}
    \label{fig:comaperOC}
\end{figure*}

\begin{table}
    \caption{Age estimates for NGC\,2281. The distribution is bimodal, with our estimate lying between the low and high groups.}
    \label{tab:ages}
    \begin{tabular}{lll}
        \hline
        \hline
        Age & Method & Reference\\
        Myr&&\\
        \hline
        250 & MS turn-off (isochrone) & \cite{1968ArA.....5....1L}\\
        280 & MS turn-off (isochrone) & \cite{1976ApJS...30..451H}\\
        300 & isochrone & \protect{\cite{1981A&A....97..235M}}\\
        610 & isochrone & \protect{\cite{2005A&A...438.1163K}}\\
        630 & NUV emission & \cite{2019AJ....158...35S}\\
        275 & isochrone & \cite{2019AJ....158..122K}\\
        526 & \ion{Ca}{ii} & \cite{2019ApJ...887...84Z}\\
        566 & H$\alpha$ &  \cite{2019ApJ...887...84Z}\\
        620 & isochrone & \protect{\cite{2020A&A...640A...1C}}\\
        630 & isochrone & \cite{2021Galax...9....7T}\\
        435 & rotation & This work\\
        \hline
    \end{tabular}
\end{table}

The two well populated open clusters NGC\,3532 (300\,Myr, \citealt{2021A&A...652A..60F}) and Praesepe (670\,Myr, \citealt{2019ApJ...879..100D}) bracket the estimated age of NGC\,2281 and our initial comparison is therefore with them. The \emph{left} panel of Fig.~\ref{fig:comaperOC} shows the colour-period diagrams for these three open clusters.
From this figure, it is immediately visible that NGC\,2281 is closer in age to NGC\,3532 than it is to Praesepe.
The slow rotator sequences of NGC\,3532 and NGC\,2281 mostly overlap, while the slow rotators in Praesepe have spun down further. The same can be seen for the fast rotators. The majority of the fast rotators in Praesepe can be found at much redder colour than in NGC\,2281. In comparison to NGC\,3532, we find very few NGC\,2281 fast rotators near the clean evolved sequence of NGC\,3532.
This suggests that NGC\,2281 is slightly older.

Fortunately, the literature provides rotation periods for two slightly older open clusters. These are \object{M\,37} (550\,Myr, \citealt{2009ApJ...691..342H}) and \object{M\,48} (450\, Myr, \citealt{2015A&A...583A..73B}).

For M\,37, we used the cleaned sample of rotation periods provided by \cite{2021ApJS..257...46G} and dereddened the photometry with the value from \cite{2009ApJ...691..342H}. Unfortunately, M\,37 is a difficult open cluster to compare with because it is distant ($d> 1$\,kpc) and suffers from differential reddening, widening all sequences in the CPD.  Nevertheless, as seen from Fig.~\ref{fig:comaperOC} the slow rotator sequences of NGC\,2281 (black) and M\,37 (green) mostly overlap, indicating a similar age. We note that \cite{2020A&A...640A...1C} assign an age of 450\,Myr to M\,37, placing it 100\,Myr younger than \cite{2009ApJ...691..342H}.

In order to obtain the cleanest set of rotation periods for M\,48, we matched the rotation periods with both \emph{Gaia} DR3 and the recent radial velocity membership study of \cite{2020AJ....159..220S} which allowed us to clean the data set by removing obvious proper motion, parallax, and radial velocity non-members. This procedure removed ten of the rotators from \cite{2015A&A...583A..73B}, and left us with 52 (mostly slow rotators) for a comparison. The photometry was dereddened with $E(B-V)=0.05$\,mag \citep{2020AJ....159..220S}. As with M\,37, the rotational distribution of M\,48 is also very similar to NGC\,2281 (Fig.~\ref{fig:comaperOC}, pink). The most recent age determination found an age of $420\pm30$\,Myr for M\,48 \citep{2020AJ....159..220S}.

From our estimates so far, NGC 2281 is in the range of $300-500$\,Myr-old. In order to obtain a narrower age range, we look at the fast rotators. It is well known that the fast rotators transition to slow rotation in a mass-dependent manner in the first Gyr of the stellar lifetime (e.g. \citealt{2003ApJ...586..464B, 2021A&A...652A..60F}). Hence, we can use these stars to achieve a higher precision in our age estimates. The evolved fast rotator sequence should be described as the lower envelope of the majority of fast rotators\footnote{A few outliers occur even in the older cluster and are often binaries \citep{2017ApJ...842...83D, 2019ApJ...879..100D} or the product of their evolution \citep{2019ApJ...881...47L}.}.

Using this criterion, we find the fast rotator distribution of NGC\,2281 to resemble the distribution of M\,37 more closely than the one of NGC\,3532. Taking into account the updated age for M\,37 (450\,Myr, \citealt{2020A&A...640A...1C}) and the overall similarities of the rotation period distributions, we conclude that all three open clusters are nearly coeval and we expect NGC\,2281 to be $435\pm50$\,Myr old, taking the mean of the ages of M\,37 and M\,48 and including the whole age uncertainty for M\,48.

In hindsight, it is not surprising that \cite{2019ApJ...887...84Z} found an age of $526-566$\,Myr from their activity analysis of NGC\,2281. One of their calibration clusters is M\,37, with an assumed age of 550\,Myr. As our rotational analysis shows, NGC\,2281 is basically coeval with M\,37 (which is now estimated to be 450\,Myr rather than 550\,Myr old). This result shows that activity and rotation are excellent and precise age tracers for intermediate aged stars.

\section{Influence of binarity on rotation and outliers to the CPD}
\label{sec:binout}
\subsection{Influence of binarity on stellar rotation}

We observe many fast rotators to be non-single stars \citep{2021A&A...652A..60F}, in keeping with the idea that binarity and fast rotation are often connected.
Binarity is usually implicated as the cause of fast rotation \citep[e.g.][]{2017ApJ...842...83D}, especially after \cite{2007ApJ...665L.155M} observed faster rotation of binaries beyond the reach of tidal interactions. We analyse the distribution of potential binaries (as defined in Sect.~\ref{sec:members}) in our rotational sample to understand whether binarity has a lasting impact on the rotation period distribution.

\begin{figure}
    \includegraphics[width=\columnwidth]{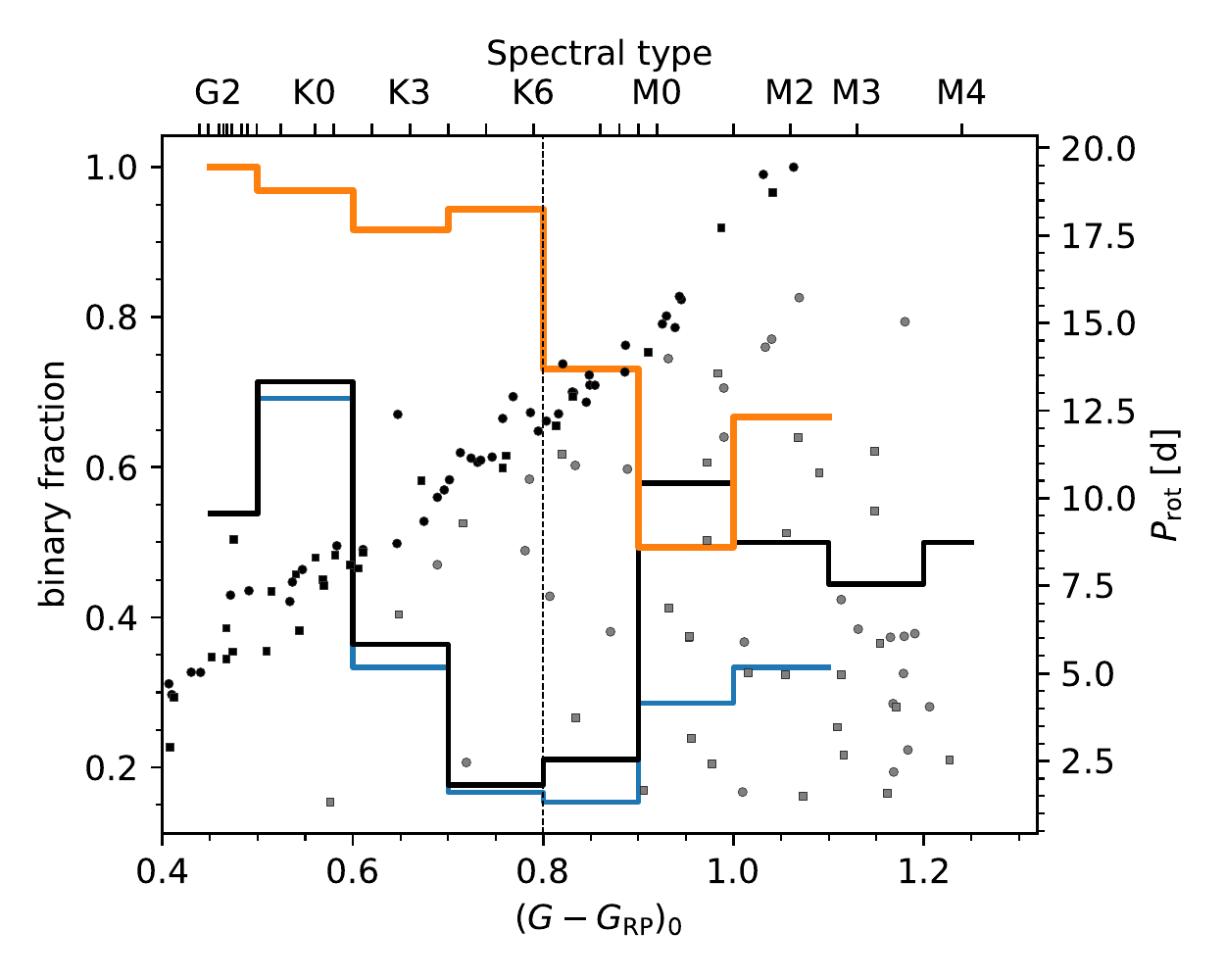}
    \caption{Fraction of binaries among the slow rotators (blue) against $(G-G_\mathrm{RP})_0$. The higher-mass stars are nearly exclusively slow rotators, hence the relative binary fraction (orange) is roughly unity. For stars beyond the threshold value (dashed line, see text), the binaries are equally distributed between fast and slow rotators, pointing to a small effect of binarity on stellar rotation. The overall binarity fraction is shown in black. In the background, we included the CPD for reference. Here, black symbols mark the slow rotators, while all other stars (fast rotators for this discussion) are in grey.}
    \label{fig:binarity}
\end{figure}

To examine this effect, we show the binarity fraction against the intrinsic colour in Fig.~\ref{fig:binarity}. From the separate binarity fractions for the slow rotators and all stars (black and blue, respectively), we find a surprisingly low binarity fraction for K stars in NGC\,2281. It can be traced back to the sparsely populated photometric binary sequence in this mass range (c.f. Fig.~\ref{fig:members}). More information on the influence of binarity on rotation can be drawn from the relative binarity fraction, that is the fraction of potential binaries on the slow rotator sequence relative to all potential binaries. For stars with $(G-G_\mathrm{RP})_0<0.8$ this fraction is of course close to unity because nearly all stars are located on the slow rotator sequence. Redwards of this threshold many faster rotators are found. However, we find the potential binaries to be equally distributed between the slow rotator sequence and the faster rotators. Hence, we find no connection between binarity and position in the CPD.

The faster rotators among stars with $(G-G_\mathrm{RP})_0<0.8$ are true outliers (see below for an individual discussion) to the rotation period distribution as we would expect all stars in this colour range to have became slow rotators by the age of NGC\,2281. These few stars may indeed be examples in which binarity has a noticeable impact on the spin evolution. We note that such fast rotators can be found in many open clusters including the aforementioned analysis of \cite{2017ApJ...842...83D}. However, the majority of the potential binaries in NGC\,2281 among the higher mass stars spin down as expected and are found on the slow rotator sequence.

In conclusion, we find that binarity is not a direct driver of fast rotation when considering all binaries equally. Of course close binaries do interact, and the outliers below the fast rotator sequence may even be tidally locked fast rotators. Nevertheless, with the large numbers of potential binaries being true slow rotators, we advocate treating (wide) binaries (i.e. binaries with $P_\mathrm{orb}>10$\,d and therefore without tidal circularization, \citealt{2005ApJ...620..970M}) the same as single stars when it comes to the overall spin evolution at this age.

\subsection{Outliers in the CPD}
\label{sec:outliers}
Upon inspection of the CPDs in Fig.~\ref{fig:cpd} and Fig.~\ref{fig:comaperOC}, certain stars can be observed not to follow the sequences. Here, we discuss the stars that we perceive as outliers individually, starting with the stars above the slow rotator sequence and moving from high to low masses.

\subsubsection{Outliers above the slow rotator sequence}
\emph{1180 [$(G-G_\mathrm{RP})_0=0.475$, $P_\mathrm{rot}=8.82$\,d]:} Although this star is only slightly above the slow rotator sequence, the period uncertainty is small enough that it is definitively above. We also find 1180 to be among the least active stars in our rotator sample, a fact expressed also in the noisy periodograms. Despite this, the chosen period is the highest consistent non-alias signal in the periodograms. With its low activity which agrees with its slow rotation, this star could be an older field star, despite all its membership criteria pointing towards cluster membership. 1180 is a photometric binary but the low variability amplitude argues against the possibility that we have observed the spot modulation of a lower-mass companion. Furthermore, we note that three other stars of similar colour are also slightly elevated on the slow rotator sequence. No other stars in this region exhibit similar behaviour.

\emph{3424 [$(G-G_\mathrm{RP})_0=0.647$, $P_\mathrm{rot}=12.4$\,d]:} This star is also above the  slow rotator sequence and its period is associated with a clear peak in the periodograms. There is some signal at $P\approx8$\,d which would be in agreement with the slow rotator sequence. However the main peak is consistent across all periodograms of the six light curves from two fields and is higher than the alias signals. Hence, we keep the period as is.

\subsubsection{Outliers below the slow rotator sequence}

In the following, we discuss stars below the slow rotator sequence and bluewards of the evolved fast rotator sequence, again moving from high to low mass.

\emph{1723 [$(G-G_\mathrm{RP})_0=0.576$, $P_\mathrm{rot}=1.33$\,d]:} This relatively blue fast rotator shows a clear rotational signal at $P_\mathrm{rot}=1.33$\,d and certain other, likely alias, periods. It also is observed to be highly active, in consonance with the short rotation period yet well above other cluster members of similar colour. In the CMD, 1723 is located slightly above the single star main sequence, indicating an unresolved lower-mass companion. It is well-known that binaries can become tidally locked. We observed one such case with a similarly short period in NGC\,2516 \citep{2020A&A...641A..51F}\footnote{We note that both Praesepe and M\,37 each host a fast rotator of similar colour and period (s. Fig.~\ref{fig:comaperOC}).}. 1723 is a bona fide member of NGC\,2281 with its proper motion and parallax near the cluster mean values.

\emph{3089 [$(G-G_\mathrm{RP})_0=0.719$, $P_\mathrm{rot}=2.46$\,d]:} Similar to 1723, this star has a very clear peak for the given period in the periodograms. However, unlike 1723, we find no evidence for binarity in the available data. We note that \emph{Gaia} lists a second star 4.2\arcsec{} from 3089. However this star is firstly, 4\,mag fainter than 3089 and secondly a distant background star. Hence, it is unlikely that we picked up a rotation modulation from the background star. In order to verify the periodicity, we use its light curve obtained by the Zwicky Transient Facility (ZTF, \citealt{2019PASP..131a8003M}) and are able to confirm this period. Further, \cite{2018AJ....156..241H} find the same period albeit with a longer period signal at 32.57\,d, too. However, we find no evidence for this period neither in our nor the ZTF light curve. In conclusion, we believe that this star is an outlier to the rotation period distribution but with the available observations, we cannot pinpoint the origin of its fast rotation.

\section{Spin down from fast to slow rotation}
\label{sec:spindown}
\subsection{Spin down time scales}
In our previous work on the 300\,Myr-old open cluster \object{NGC\,3532} \citep{2021A&A...652A..60F}, we investigated the spin down from fast to slow rotation of early K\,dwarfs. With three older, coeval open clusters, we can now roughly estimate the spin down of later-type stars and even probe the spin down of the least massive fast rotators by comparing NGC\,2281 with the older Praesepe open cluster \citep{2019ApJ...879..100D}.

\begin{figure}
    \includegraphics[width=\columnwidth]{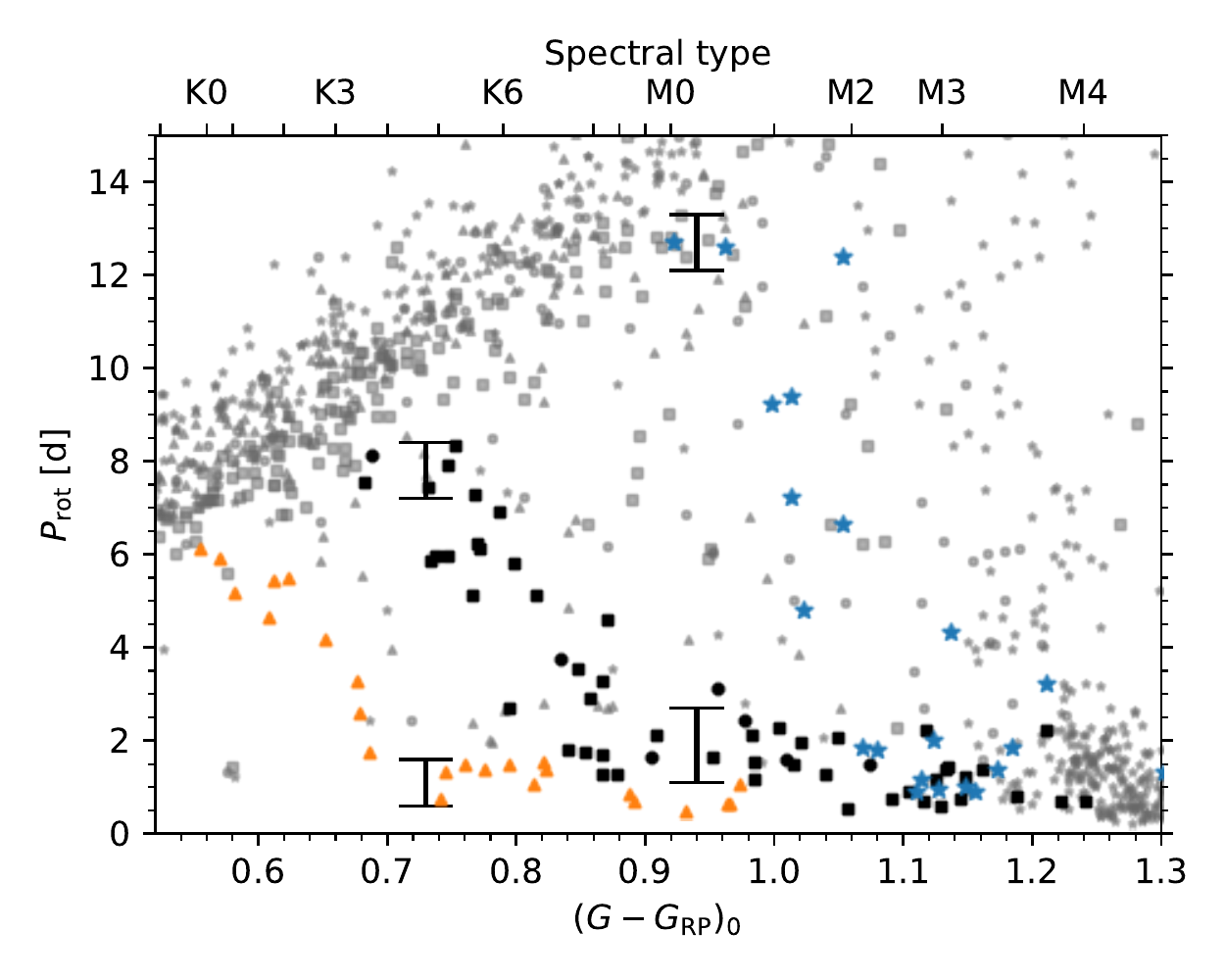}
    \caption{Evolution of the fast rotator sequence in time. The highlighted sequences (from left to right) belong to NGC\,3532 (300\,Myr, orange triangles), NGC\,2281 (430\,Myr, black circles), M\,37 (450\,Myr, black squares), M\,44 (Praesepe 670\,Myr, blue stars). The large errorbars indicate the positions and ranges of rotation periods used in the text to estimate the spin-down rate. Grey symbols in the background denote other rotators in these open clusters and are marked with the respective symbols.}
    \label{fig:fastev}
\end{figure}

To facilitate the process of estimating the spin down rate, we highlight the evolved fast rotator sequences of NGC\,3532 (300\,Myr), NGC\,2281, M\,37 ($\sim${}440\,Myr), and \object{M\,44} (Praesepe, 670\,Myr) in Fig.~\ref{fig:fastev}. The evolutionary sequence is clearly visible as lower mass stars evolve further in the older open clusters, while the higher mass stars have joined the slow rotator sequence.

In Fig.~\ref{fig:fastev}, we have marked approximate rotation period ranges for two exemplary spectral types, namely mid-K and early-M. These two masses were selected because they mark the ends of the (un)evolved fast rotator sequences in the open clusters. Namely the mid-K ($(G-G_\mathrm{RP})_0=0.73$) stars mark the most-massive unevolved fast rotators in NGC\,3532, while the early-M ($(G-G_\mathrm{RP})_0=0.94$) stars are chosen as the most-massive evolved fast rotators in Praesepe. We than selected the appropriate stars in the combines data set of NGC\,2281 and M\,37. The mid point and ranges in periods are estimates, which is sufficient for this kind of calculation.

For the mid-K stars, we find $\Delta P_\mathrm{rot}=6.7\pm1.1$\,d which results in a spin down of $\frac{\Delta P}{\Delta t}=0.048\pm0.08$\,d\,Myr$^{-1}$. Assuming the above determined age of NGC\,2281 and M\,37 to be $t\approx 440$\,Myr. For early-M\,dwarfs, we find $\Delta P_\mathrm{rot}=10.85\pm1.35$\,d and $\frac{\Delta P}{\Delta t}=0.047\pm0.06$\,d\,Myr$^{-1}$, when assuming 670\,Myr for the age of M\,44 as \cite{2019ApJ...879..100D} do.

We note that the assumed age difference is uncertain not only due to the uncertain age of M\,44 but due to the stellar mass chosen for the spin down estimate. Due to the large age difference, we were not able to use fast rotators at the bottom of the evolved fast rotator sequence but had to use slightly redder stars that have not yet started transitioning. Based on their position in the CPD, we expect these stars to evolve off the fast rotators sequence in a few Myr. Yet, $\Delta t$ could be smaller, increasing the spin down rate. Without additional data in between the ages of NGC\,2281 and M\,44 the estimated spin down rate should be handled with care.

Within the uncertainties of estimates, we find mid-K and early-M stars to spin down with the same rate. However, the value is different from our estimate of early-K stars in NGC\,3532 \citep{2021A&A...652A..60F}. There, we found $\frac{\Delta P}{\Delta t}=0.063$\,d\,Myr$^{-1}$. This discrepancy leads us to suspect that the spin down from fast to slow rotation of early-K stars might be influenced by an additional mass dependence. In recent years, it has become clear that the spin down of K-type slow rotators significantly deviates from earlier-type stars (e.g. \citealt{2018ApJ...862...33A, 2019ApJ...879...49C, 2020A&A...644A..16G}). A possible solution was proposed by \cite{2020A&A...636A..76S} with the (re-)introduction of the core-envelope coupling time \citep{1991ApJ...376..204M, 1993MNRAS.261..766J}. More investigations are needed to understand whether this time scale also has an impact on the spin down from fast to slow rotation.

\subsection{Towards an empirical description of the spin down from fast to slow rotation}

\begin{figure}
    \includegraphics[width=\columnwidth]{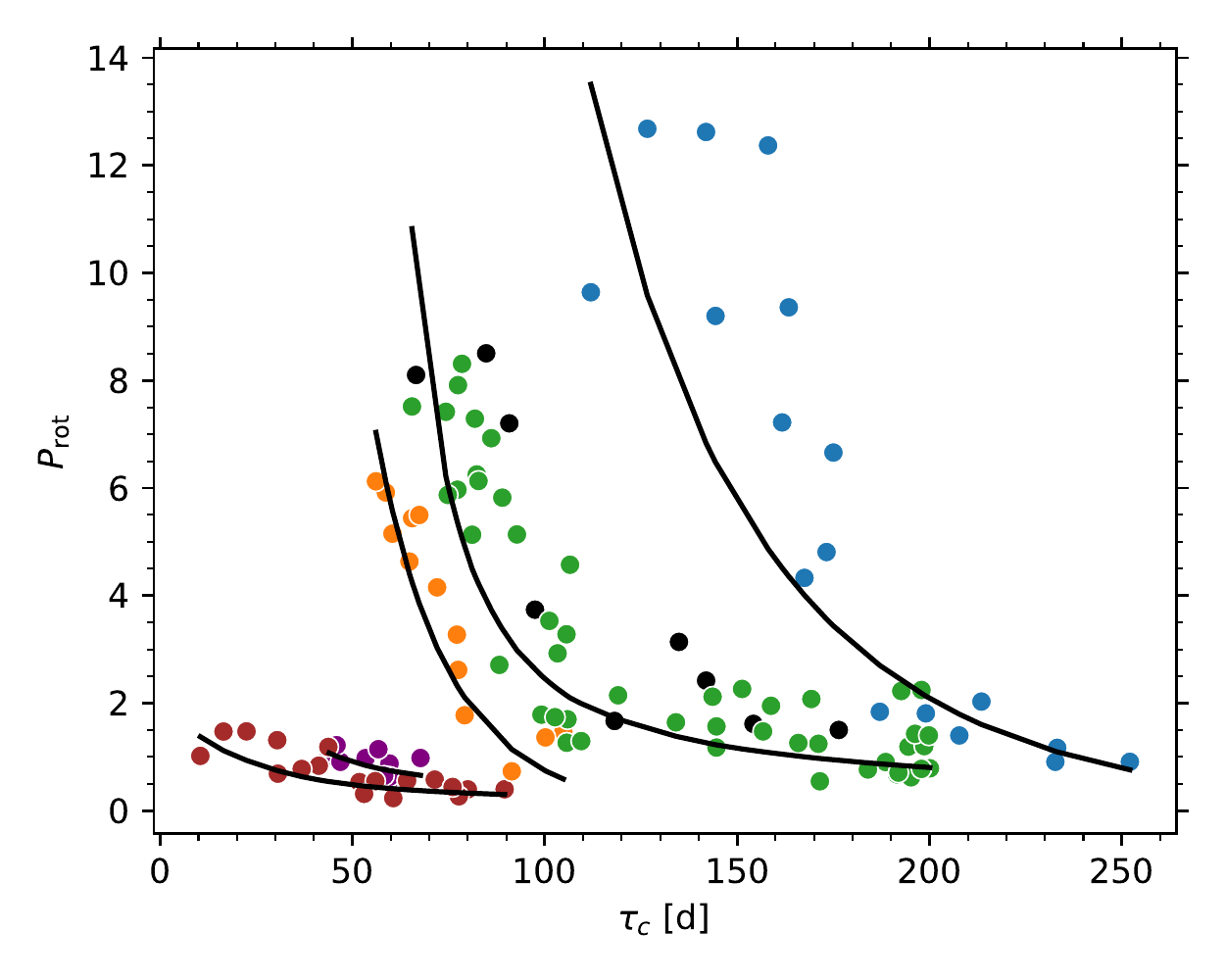}
    \caption{Rotation periods against convective turnover time ($\tau_c$) for fast rotators selected from several young open clusters. The sequences show a clear exponential dependence on $\tau_c$. NGC\,2281 and M\,37 are combined into a single fit. The black lines show cluster-by-cluster exponential fits to the data. The individual open clusters are Pleiades (brown, \citealt{2016AJ....152..113R}), NGC\,2516 (purple, \citealt{2020A&A...641A..51F}), NGC\,3532 (orange, \citealt{2021A&A...652A..60F}), NGC\,2281 (black, this work), M37 (green, \citealt{2009ApJ...691..342H}), and Praesepe (blue, \citealt{2019ApJ...879..100D}).}
    \label{fig:fastfits}
\end{figure}

With the transition from fast to slow rotation being recognized as a well ordered evolutionary process, we aim to understand its dependence on the stellar parameters. As shown in our previous work \citep{2010ApJ...722..222B, 2020A&A...641A..51F}, the rotation periods of young open clusters can be parametrized with the convective turnover time as its main mass dependence. In Fig.~\ref{fig:fastfits}, we show only the fast rotators from several young open clusters against their convective turnover time. The shape immediately suggests an exponential dependence as originally proposed in \cite{2003ApJ...586..464B}. For the time being, we show individual exponential dependencies for each open cluster (lines in Fig.~\ref{fig:fastfits}). In future work, we will aim to provide a unifying solution which would enable an accurate age determination from fast rotators in open clusters and other coeval stellar groups with ages in between that of the Pleiades and Hyades.

\section{Conclusions}
\label{sec:conclusions}
Cool star rotation periods are becoming an increasingly important tool in obtaining ages for open clusters. In this work, we present new rotation periods measurements for cool star members of the open cluster NGC\,2281 to derive a more precise age of 435$\pm$50\,Myr from gyrochronology in comparison with the previously literature values which range from 250\,Myr to 630\,Myr. Further, we provide new data to understand the transition from fast to slow rotation in more detail.

The rotation periods were measured from an eight month-long photometric time series obtained with the robotic STELLA telescope.  With the combination of three filters and exposure times, we were able to measure rotation periods for \ngoodper{} of the \ngoodlc{} cool star members included in our time series photometry. This large fraction (70\,\%) allows us to construct a rich colour-period diagram for NGC\,2281.

The CPD shows a clear slow rotator sequence stretching from late-F to early-M\,dwarfs with few outliers above the sequence. This clarity is expected for coeval populations but it was especially enabled by the very accurate membership determination made possible by \emph{Gaia}. Outliers among the bluer stars are likely tidally interacting binary stars, but further evidence of binarity impacting the spin down has not been found.

The photometric activity of stars in NGC\,2281 follows the shape expected from prior observations. It allows us to identify and suppress likely alias periods.

We determined the rotational age of NGC\,2281 by comparing its rotation period distribution with those of the open clusters M\,37 and M\,48. Recent work on these two clusters has assigned very similar ages and we are able to confirm with the cool star rotation periods that NGC\,2281, M\,37, and M\,48 are indeed essentially coeval. We estimate NGC\,2281 to be $435\pm50$\,Myr old, which places it neither in the low age group of prior ages estimates, nor in the high-age group, but intermediate between these determinations.

The additional open cluster between the ages of NGC\,3532 (300\,Myr) and the two $600-800$\,Myr old clusters Praesepe and Hyades, together with the more confined age of M\,37 enabled us to advance the empirical description of the spin down from fast to slow rotation. We find that the spin down is stronger for early-K stars than for mid-K and early-M stars, leading to the possibility of an additional mass dependence as seen for K\,dwarf slow rotators.

We find that the evolution of fast rotators in open clusters can indeed be described in terms of an exponential dependence of the convective turnover time. With this insight, future work may enable accurate age determinations from fast rotator sequences of young open clusters.

In conclusion, our work fills crucial gaps in the empirical picture of the spin down from fast to slow rotation. The rotation periods for NGC\,2281 and their comparison with other open clusters again demonstrates how powerful cool star rotation periods are in determining open cluster ages.

\begin{acknowledgements}
    We are grateful to the anonymous referee for providing comments that improved the paper.
    STELLA (stella.aip.de) was made possible by funding through the State of Brandenburg (MWFK) and the German Federal Ministry of Education and Research (BMBF). The STELLA facility is a collaboration of the AIP in Brandenburg with the IAC in Tenerife.
    This research has made use of NASA's Astrophysics Data System Bibliographic Services and of the SIMBAD database and the VizieR catalogue access tool, operated at CDS, Strasbourg, France.
    The Digitized Sky Survey was produced at the Space Telescope Science Institute under U.S. Government grant NAG W-2166. The images of the Digitized Sky Survey are based on photographic data obtained using the Oschin Schmidt Telescope on Palomar Mountain and the UK Schmidt Telescope.
    This publication makes use of data products from the Wide-field Infrared Survey Explorer, which is a joint project of the University of California, Los Angeles, and the Jet Propulsion Laboratory/California Institute of Technology, and NEOWISE, which is a project of the Jet Propulsion Laboratory/California Institute of Technology. WISE and NEOWISE are funded by the National Aeronautics and Space Administration.
    This work has made use of data from the European Space Agency (ESA) mission \emph{Gaia} (\url{https://www.cosmos.esa.int/gaia}), processed by the \emph{Gaia} Data Processing and Analysis Consortium (DPAC, \url{https://www.cosmos.esa.int/web/gaia/dpac/consortium}). Funding for the DPAC has been provided by national institutions, in particular the institutions participating in the \emph{Gaia} Multilateral Agreement.
    Based on observations obtained with the Samuel Oschin Telescope 48-inch and the 60-inch Telescope at the Palomar Observatory as part of the Zwicky Transient Facility project. ZTF is supported by the National Science Foundation under Grant No. AST-2034437 and a collaboration including Caltech, IPAC, the Weizmann Institute for Science, the Oskar Klein Center at Stockholm University, the University of Maryland, Deutsches Elektronen-Synchrotron and Humboldt University, the TANGO Consortium of Taiwan, the University of Wisconsin at Milwaukee, Trinity College Dublin, Lawrence Livermore National Laboratories, and IN2P3, France. Operations are conducted by COO, IPAC, and UW. The ZTF forced-photometry service was funded under the Heising-Simons Foundation grant \#12540303 (PI: Graham).
    \newline
    \textbf{Software:}
    This research made use of \textsc{Astropy}, a community-developed core Python package for Astronomy \citep{2013A&A...558A..33A}.
    This work made use of \textsc{Topcat} \citep{2005ASPC..347...29T}.
    This research made use of the following \textsc{Python} packages:
    \textsc{Pandas} \citep{pandas};
    \textsc{NumPy} \citep{numpy};
    \textsc{MatPlotLib} \citep{Hunter:2007};
    \textsc{IPython} \citep{ipython};
    \textsc{SciPy} \citep{scipy};
    \textsc{seaborn} \citep{Waskom2021}

\end{acknowledgements}

\bibliographystyle{aa} 
\bibliography{N2281.bib} 

\begin{appendix}

\section{Photometric membership}
\label{app:members}

In order to include stars below the faint limit of the membership definition of \cite{2019AJ....158..122K}, we define additional photometric members from \emph{Gaia} and WISE photometry. We use a $(G-G_\mathrm{RP})_0$ CMD and a [$(G-G_\mathrm{RP})_0$, $(W1-W2)_0$] colour-colour diagram. Only stars fulfilling both photometric criteria below are considered members.

To compare the observed colours to isochrones, we dereddened the photometry for all stars in the field with the cluster value $E(B-V)=0.11$ using the reddening coefficients from \cite{2018MNRAS.479L.102C} (\emph{Gaia}) and \cite{2013MNRAS.430.2188Y} (WISE). Photometric members are stars within $\Delta(G-G_\mathrm{RP})_0=0.03$\,mag and $\Delta(W1-W2)_0=0.05$\,mag from the cluster sequence in the respective diagrams. We use the empirical sequence from \cite{2013ApJS..208....9P} for the \emph{Gaia} CMD) and the theoretical isochrone from the Dartmouth Stellar Evolution Database \citep{2008ApJS..178...89D} for the \emph{Gaia}-WISE colour-colour diagram, due to missing data in the former.

We define photometric members as stars which typically satisfy both photometric criteria. An exception is the case in which one criterium cannot be fulfilled due to missing data. In these cases a photometric member may satisfy only one criterium.

\begin{figure}
    \includegraphics[width=\columnwidth]{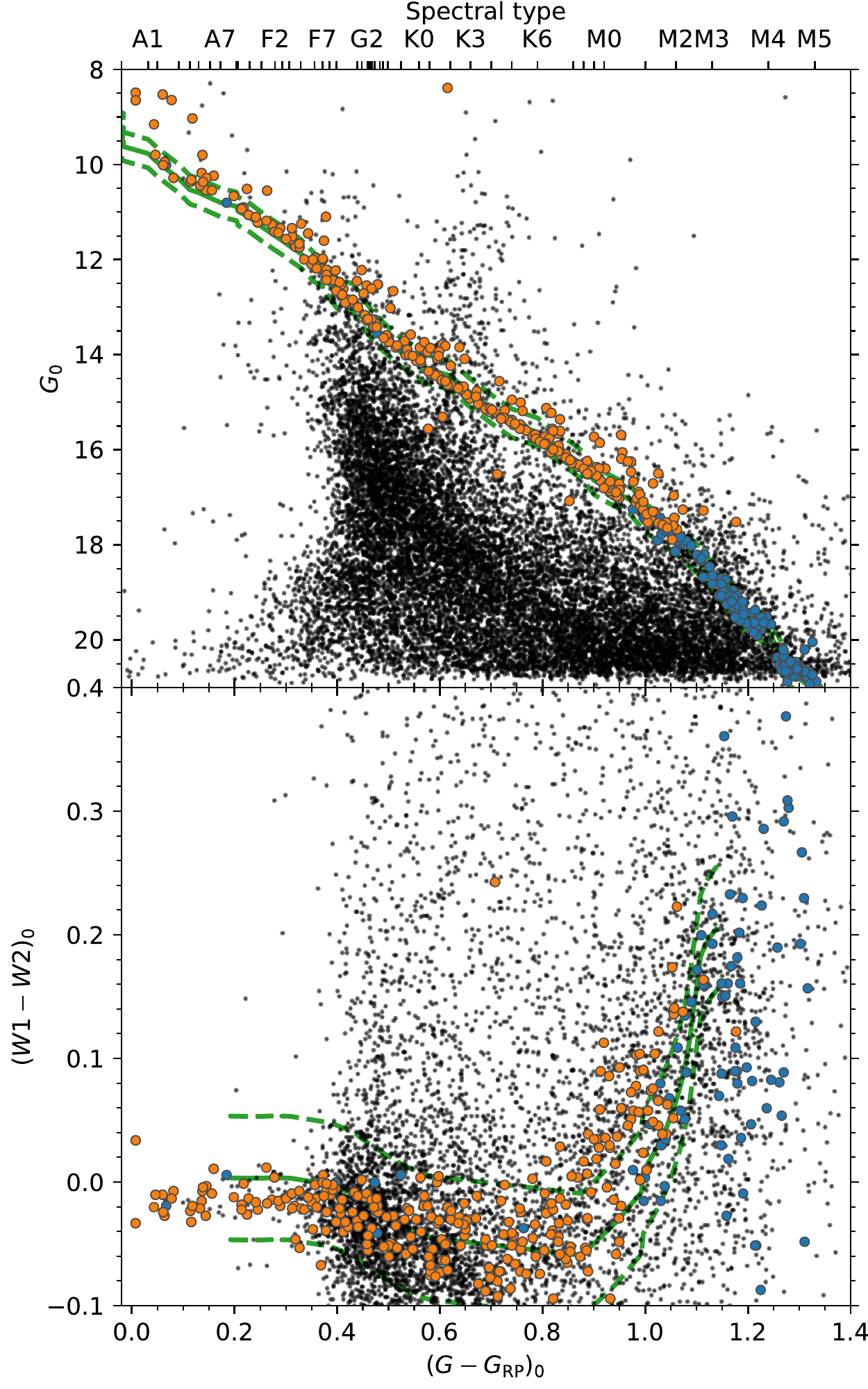}
    \caption{Photometric membership definition for NGC\,2281. Members from  \cite{2019AJ....158..122K} are marked in orange and additional members from our analysis in blue. The solid green line gives the single star isochrone and the dashed lines enclose the region in which we considered stars as members. Field stars are shown in black.}
    \label{fig:phmem}
\end{figure}

Fig.~\ref{fig:phmem} shows all members together with the limits set by our photometric criteria. Only stars fulfilling both photometric criteria and the proper motion criterium (Sect.~\ref{sec:members}) are considered members and shown here.

We note that our cluster sequence of the final members has a gap among the low mass stars in the \emph{Gaia} CMD. The position of this gap coincides with the transition from partially to fully convective stars at M\,3.5 (see also \citealt{2018ApJ...861L..11J}).

\end{appendix}

\end{document}